%Paper: hep-th/9406021
%From: ejw@cuphyf.phys.columbia.edu (Erick J. Weinberg)
%Date: Fri, 3 Jun 1994 16:53:44 -0400

\input phyzzx

\def\la{\mathrel{\mathpalette\fun <}}
\def\ga{\mathrel{\mathpalette\fun >}}
\def\fun#1#2{\lower3.6pt\vbox{\baselineskip0pt\lineskip.9pt
  \ialign{$\mathsurround=0pt#1\hfil##\hfil$\crcr#2\crcr\sim\crcr}}}

\def\pr#1#2#3#4{Phys. Rev. D{\bf #1}, #2 (19#3#4)}
\def\np#1#2#3#4{Nucl. Phys. {\bf B#1}, #2 (19#3#4)}

\nopagenumbers
\line{\hfil CU-TP-634}
\line{\hfil hep-th 9406021}
\vglue .5in
\centerline {\twelvebf  NONTOPOLOGICAL MAGNETIC MONOPOLES AND}
\centerline {\twelvebf  NEW MAGNETICALLY CHARGED BLACK HOLES }
\vskip .3in
\centerline{\it Kimyeong Lee and Erick J. Weinberg}
\vskip .1in
\centerline{Physics Department, Columbia University}
\centerline{New York, New York 10027}
\vskip .4in
\baselineskip=14pt
\overfullrule=0pt
\centerline {\bf Abstract}
\medskip
\centerline{\vbox{\hsize=5truein
    The existence of nonsingular classical magnetic monopole solutions
is usually understood in terms of topologically nontrivial Higgs field
configurations.  We show that finite energy magnetic monopole
solutions also exist within a class of purely Abelian gauge theories
containing charged vector mesons, even though the possibility of
nontrivial topology does not even arise.  provided that certain
relationships among the parameters of the theory are satisfied.  These
solutions are singular if these relationships do not hold, but even
then become meaningful once the theory is coupled to gravity, for they
then give rise to an interesting new class of magnetically charged
black holes with hair. }}

\baselineskip=18pt
\vskip .1in
\footnote{}{\twelvepoint This work was supported in part by
the National Science Foundation Presidential Young Investigator
Program, the Alfred P. Sloan Foundation, and the US Department of
Energy}

\pagenumbers

      It was shown by 't Hooft and Polyakov\ref{ G. 't Hooft,
\np{79}{276}74; A.M. Polyakov, Pis'ma Zh.
Eksp. Teor. Fiz. {\bf 20}, 430 (1974) [JETP Lett. {\bf 20}, 194
(1974)].} that classical finite energy
magnetic monopole solutions occur in certain spontaneously broken
non-Abelian gauge theories.   The existence of these solutions is often
understood in terms of topologically nontrivial
Higgs field configurations.   In this letter we show that finite energy
magnetic monopoles can also be obtained in a class of purely Abelian
theories in which such topological consideratations do not even
arise, provided that certain relationships among the parameters of the
theory are satisfied.   These solutions are singular if these
relationships do not hold, but even then become meaningful once the
theory is coupled
to gravity, for they then give rise to a new class of magnetically charged
black holes with hair.

       To illustrate these ideas, we consider a theory with
electromagnetism coupled to a charged vector field $W_\mu$ and a
neutral scalar field $\phi$.  The spin-1 $W$-particles have electric
charge $e$ and a magnetic moment $ g d_{\mu\nu} \equiv ieg(W_\mu^*
W_\nu - W_\nu^* W_\mu) $ with $g$ assumed to be
positive\ref{If $g$ is negative, finite energy flat
spacetime monopole solutions such as we describe do not exist.
However, our results concerning new black hole solutions remain true
with only minor modifications.}. They have a $\phi$-dependent mass
$m(\phi)$ that takes on a nonzero value $m_W = m(v)$ when the scalar
field takes on its vacuum value $v$ but vanishes at some other value
of $\phi$, which we arbitrarily choose to be $\phi=0$.  Adding a
quartic $W$ self-coupling proportional to $d_{\mu\nu}^2$ (other
interactions are also possible), we obtain the Lagrangian
$$ \eqalign{ {\cal L} &=  -{1\over 2}|D_\mu W_\nu
- D_\nu W_\mu|^2
 -{1\over 4} F_{\mu\nu}F^{\mu\nu} + {g\over 4} d_{\mu\nu} F^{\mu\nu}
  -{\lambda\over 4} d_{\mu\nu} d^{\mu\nu} \cr
  &\qquad
  + m^2(\phi) |W_\mu|^2 + {1\over 2} \partial_\mu \phi \partial^\mu\phi
   - V(\phi) \cr}
    \eqn\lagrangian$$
where $F_{\mu\nu} = \partial_\mu A_\nu - \partial_\nu A_\mu$ is the
electromagnetic field strength and
 $D_\mu W_\nu = (\partial_\mu -ie A_\mu) W_\nu$ is the $U(1)$ covariant
derivative.
If $g=2$, $\lambda= 1$,
and $m(\phi)= e\phi$, this is in fact the unitary gauge form of the
Lagrangian for an
$SU(2)$ gauge theory spontaneously broken to $U(1)$ by a triplet Higgs
field.   Similarly, for  $g=2$, $\lambda= (\sin \theta_W)^{-2}$,
and $m(\phi)= e\phi/2$ we have the unitary gauge form of the standard
electroweak theory, but with all terms involving the $Z$ or fermions
omitted.  However, for generic values of $g$ and $\lambda$
an extension to a non-Abelian symmetry is not
possible\ref{Apart from special cases such as those
noted here, the theory described by Eq.~\lagrangian\ is
nonrenormalizable. This will not affect the analysis of this paper,
which is largely classical.}.

      It is useful to display the energy density corresponding to this
Lagrangian.  For static configurations with $A_0 = W_0 =0$, this may be
written as
$$ \eqalign{ {\cal E} &=
   {1\over 4}\left( 1- {g^2 \over 4\lambda}\right) F_{ij}^2
   + {g^2 \over 16\lambda}
     \left( F_{ij} - {2\lambda \over g} d_{ij} \right)^2
    +{1\over 2} |D_iW_j - D_jW_i|^2
   \cr &\qquad + m^2(\phi) |W_i|^2  +{1\over 2} (\partial_i \phi)^2
        +  V(\phi) \,.}
   \eqn\energy $$
In any magnetically charged configuration, the magnetic field will be
at least as
singular as $1/r^2$ at the position of the magnetic charge.   We will show
below that in certain situations
$W_i$ can be chosen so as to cancel the singularity in the second
term on the right hand side of this equation.  Having done this, we are
still left with a $1/r^4$ singularity from the first term.  This leads to
three cases:  (1) If $g^2 > 4\lambda$, there are monopole configurations with
negative infinite energy.  The vacuum is therefore unstable against
production of monopole-antimonopole pairs and the theory must be discarded.
(2)  If $g^2 < 4\lambda$, all monopole solutions have
positive infinite energy.  We will return to this case later.   (3)  If $g^2
= 4\lambda$, the first term on the right hand side of Eq.~\energy\ is
absent, and finite energy monopole solutions exist, as we now demonstrate.

      To begin, we recall that a point magnetic charge gives rise to a
radial magnetic field of magntitude $Q_M/ r^2$ that is derived
from a vector potential that necessarily possesses a
Dirac string singularity along some line running from the monopole to
spatial infinity.   In the quantum theory, Dirac strings are acceptable as
long as they cannot be detected through the Aharanov-Bohm effect by particles
encircling the string.  The analogous criterion in the classical theory is
that the string not be detectable through the interference of waves in the
$W$ field (or any other charged field) passing on either side of the string.
In both cases, this leads to the quantization condition $Q_M = q/e$ where
$q$ is either an integer or a half-integer and $e$ is the smallest
electric charge in the
theory\ref{To understand how the same quantization
condition can arise in the classical and quantum theories, it is
helpful to work in units where $\hbar$ and $c$ have not been set equal
to unity.   To give $e$ the dimensions appropriate to the electric
charge of a
particle, the covariant derivative must be written in the form
$D_\mu = \partial_\mu -i(ec/\hbar) A_\mu$.  The classical quantization
condition is then that $ecQ_M/\hbar$ be an integer or half-integer,
which is the same as the usual quantum mechanical result.}.
We will concentrate for the present on the case $Q_M = 1/e$.  This has
the advantage of allowing spherically symmetric $W$ fields, which
cannot occur\ref{A.H.~Guth and E.J.~Weinberg, Phys. Rev.~D{\bf 14},
1660 (1976); E.J. Weinberg, Phys. Rev.~D{\bf 49},
(1994).}
for any other value of $Q_M$, and will also allow us to
make the connection with the 't~Hooft-Polyakov solution.

      The electromagnetic vector potential for the unit charged point
monopole may be written as
$$  A_i = - \epsilon_{ij3}\, {\hat r}_j {1\over er}
     \left({ 1-\cos\theta \over \sin^2\theta }\right) \,.
\eqn\Dirac $$
This is spherically symmetric in the sense that the effects
of a spatial rotation can be compensated by a gauge transformation.  Any
charged vector field that is also invariant under the same combination of
rotation and gauge transformation can be written in the form
$$  \eqalign  {  W_x & = -{i\over \sqrt{2}} {u(r)\over er}
   \left[ 1 - e^{i\phi} \cos\phi (1-\cos \theta) \right]    \cr
                 W_y  & = {1\over \sqrt{2}} {u(r)\over er}
   \left[ 1 + i e^{i\phi} \cos\phi (1-\cos \theta) \right]     \cr
                 W_z  & = {i\over \sqrt{2}} {u(r)\over er}
                              e^{i\phi} \sin\phi \,. }
\eqn\Wansatz $$
The singularities of these fields along the negative $z$-axis are purely
gauge artifacts, and can be removed by a gauge transformation that moves the
Dirac string.   The singularity at the origin cannot be removed by a gauge
transformation but, as we shall see, it does not entail any singularity in
the energy density.

   The vector field \Wansatz\ leads to a purely radial magnetic moment
of magnitude $- {|u|^2 /  er^2}$
By setting $u(0) = \sqrt{g/2\lambda}=1 $, the
$1/r^2$ singularities of $F_{ij}$ and
$d_{ij}$ can be made to cancel in the energy
density.  This leaves two other potentially singular
contributions.  The most dangerous is the term containing the
covariant curl, $D_iW_j - D_jW_i$, in which one might expect
a $1/r^4$ singularity in the energy density to arise from the
angular derivatives.
However, explicit calculation reveals that the contributions
from these angular derivatives cancel, leaving only a term
proportional to $(u'/r)^2$ that causes no problem as long as
$u(r) - u(0)$ is of order $r^2$.  The mass term could also give a
singular energy density, proportional to $1/r^2$, but this can be avoided
by requiring that $\phi(0)=0$, so that $m(\phi)$ vanishes at the
origin.  This shielding of the magnetic charge is energetically favorable
only out to a distance $R_{\rm mon} \sim \sqrt{g}\, m^{-1}_W$; beyond
this distance
the energy is minimized if $u$ and $\phi$ rapidly approach $0$ and
$v$, respectively.  Standard arguments then show that
configuration of minimum energy (which is a
solution to the field equations everywhere except, possibly, at the
origin) has a total energy $ M_{\rm mon} \sim {m_W / e^2\sqrt{g}}$.

     It was noted above that our Lagrangian is equivalent to an $SU(2)$ model
if $g=2$ and $\lambda=1$.   In this case one
can verify that the solution we have found
is simply the familiar $SU(2)$ monopole solution
$$ \eqalign { V_j^a &= \epsilon_{jak}   \hat r_k {1-u(r) \over er} \cr
     \phi^a &= h(r) \hat r^a }
        \eqno\eq $$
transformed into a gauge where the Higgs field has a constant
direction in internal space.

     Let us now return to the case
$g^2< 4\lambda$.  Although the first term on the right hand side of
Eq.~\energy\ gives a divergent contribution to the energy, the
remaining contributions can be made finite by the choice
$u(r) = \sqrt{g/2\lambda} + O(r^2)$,
$\phi(r) = O(r)$ near the origin.  This leads to an energy density which at
short distances is essentially that of a point monopole with a reduced
charge $ Q_{eff} = [1 - (g^2/ 4\lambda)]^{1/2} \, Q_M$.
As with the previous case, $u$ vanishes rapidly for
$r > R_{\rm mon} \sim \sqrt{g}\,m^{-1}_W$; in this region
the energy density is simply that of an ordinary unit charged point
monopole.

    While infinite energy solutions such as this do not correspond to
particles of the theory \lagrangian, they acquire physical
significance when the
theory is coupled to gravity, since the singularity can then be hidden
behind the event horizon of a black hole.  Let us begin by recalling
the Reissner-Nordstrom solutions, which describe charged black holes
in a theory governed by the coupled Einstein-Maxwell equations.
For magnetic charge $Q_M=q/e$ the metric is
$$  ds^2 = B(r)dt^2 - A(r) dr^2 - r^2 d\theta^2 - r^2 \sin^2\theta\, d\phi^2
      \eqn\metric $$
where
$$ B = A^{-1} = 1 - {2MG \over r} + {4\pi G q^2 \over e^2r^2}
    \eqno\eq $$
while the vector potential is precisely the Dirac monopole potential of
Eq.~\Dirac.  Solutions with horizons exist for all values of the black hole
mass greater than the extremal mass
$$ M^{RN}_{\rm ext}(q) = {\sqrt{ 4\pi} |q|\over e}\,m_{\rm Pl}
     \eqno\eq $$
(the Planck mass $m_{\rm Pl} = G^{-1/2}$), for which the horizon radius
$r_H$ takes its minimum value,
$$ r^{RN}_{\rm ext}(q) =  {\sqrt{ 4\pi} |q|\over e} \,m_{\rm Pl}^{-1} \,.
      \eqno\eq$$

     The Reissner-Nordstrom black hole is also a solution to the
spontaneously broken $SU(2)$ theory, provided that the massive vector
field vanishes and the scalar field takes on its vacuum value
everywhere.  However, there can also be magnetically charged black
holes with ``hair'', i.e., nontrivial $W$ and $\phi$ fields outside
the horizon.  Such solutions\Ref\hair{K. Lee, V.P. Nair and E.J. Weinberg,
Phys. Rev. D{\bf 45}, 2751 (1992);
P. Breitenlohner, P. Forgacs and D. Maison, \np{383}{357}92.}
exist for a range of parameters that
roughly corresponds to the condition that the ``Schwarzschild radius''
$2MG$ be less than $R_{\rm mon}$.  At large distances these approach the
Reissner-Nordstrom solution, while for $r \ll R_{\rm mon}$ the
metric is approximately Schwarzschild.  Thus, these solutions are most
naturally viewed as 't~Hooft-Polyakov monopoles with small
Schwarzschild black holes located at their centers.  In particular,
there is no minimum value for the horizon radius, and hence no
extremal solution.

     The existence of these solutions can be understood by considering
small fluctuations about a Reissner-Nordstrom solution with magnetic
charge $1/e$.  If the horizon distance $r_H \la R_{\rm mon}$, the energy
density just outside the horizon can be lowered by the creation of a
$W$ field with its magnetic moment arranged to shield the Coulomb
field.  One can show\ref{ K. Lee, V.P. Nair and E.J. Weinberg, Phys.
Rev. Lett. {\bf 68}, 1100 (1992).}
that the Reissner-Nordstrom
solution becomes
unstable for $r_H$ less than a critical value of order $R_{\rm mon}$; the
configuration to which this instability leads is just such a black
hole with hair.

      Since these arguments do not
depend on the existence of an underlying $SU(2)$ symmetry, there
should also be nontrivial solutions for the more general Lagrangian of
Eq.~\lagrangian.   Let us examine this possibility more
closely.  We assume static spherically symmetric matter fields
as in Eqs.~\Dirac\ and \Wansatz, and write the metric as in
Eq.~\metric.  The matter
portion of the action then takes the form
$$ S_{\rm matter} = -4\pi\int dt\, dr\, r^2\sqrt{AB}
     \left[ {K(u,\phi) \over A} + U_1(u,\phi)
     + {1\over 2e^2r^4}\left(1 - {g^2\over 4\lambda}\right)\right]
    \eqno\eq$$
where
$$ K = {{u^\prime}^2 \over e^2r^2} + {1\over 2} {\phi^\prime}^2
     \eqno\eq$$
$$  U_1 = {\lambda\over 2e^2r^4}
    \left( u^2- {g\over 2\lambda} \right)^2
        + {u^2 m^2(\phi) \over r^2 } + V(\phi) \,.
     \eqno\eq$$
(Primes denote differentiation with respect to $r$.)
If we define a function $F(r)$ by
$$ A(r) = \left[ 1 -{2G {F}(r)\over r} +{4\pi G\over r^2 e^2}
     \left( 1-{ g^2 \over 4\lambda} \right) \right]^{-1}
     \eqn\Ainv $$
the gravitational field equations imply that
$$ F^\prime = 4\pi r^2 \left( {K\over A} + U_1  \right) \,.
      \eqn\Meq$$

    The black hole mass $M = F(\infty)$.  A lower bound on this
can be obtained by noting
that the horizon radius $r_H$ is a zero of
$A(r)^{-1}$ and that Eq.~\Ainv\ shows that such zeroes can exist only
if
$$   F(r_H) \ge {\sqrt{4\pi}\over e} m_{\rm Pl}
    \left( 1-{ g^2 \over 4\lambda} \right)^{1/2}
  = M^{RN}_{\rm ext} \left( 1-{ g^2 \over 4\lambda} \right)^{1/2} \,.
     \eqno\eq  $$
The total mass exceeds $F(r_H)$ by an amount equal to the integral
of Eq.~\Meq\ from $r_H$ to $\infty$.  With $u(r)$ behaving as
we expect, this integral will be roughly equal to $M_{\rm mon}$.
Because
a nontrivial $u(r)$ is energetically favorable only out to distances
of order $R_{\rm mon} \sim \sqrt{g}\,m_W^{-1}$,
solutions of the type we seek should exist only if $r_H \la R_{\rm mon}$.
This gives an upper bound on the mass, and implies that
$$  M^{RN}_{\rm ext} \left( 1-{ g^2 \over 4\lambda} \right)^{1/2}  +
M_{\rm mon}
\,\la\, M \,\la \, {\sqrt{g} \, m_{\rm Pl}^2 \over 2 m_W} + M_{\rm mon}
   \eqn\extmass $$
while the horizon radius obeys
$$    r^{RN}_{\rm ext}  \left( 1-{ g^2 \over 4\lambda} \right)^{1/2} \le
    r_H  \la R_{\rm mon} \,.
   \eqn\extrad $$
Examining the lower bounds, we see that if $g^2 \ne 4\lambda$ (i.e.,
if there are no finite energy monopoles) there is a new type of
extremal black hole with horizon distance and mass both less than
those of the extremal Reissner-Nordstrom black hole.

     Let us now consider solutions with $q \ne 1$.  Except for the
singular point monopole in flat spacetime or the Reissner-Nordstrom
black hole, none of these can be spherically symmetric and analysis
of the field equations becomes much more difficult.
However, another line of attack is available.  Consider first the
case $g^2 < 4\lambda$, where we know that only black hole solutions
are possible.
The Reissner-Nordstrom solutions are classically unstable if it is
energetically favorable to shield the magnetic charge by creating a
cloud of $W$ particles just outside the horizon.  For $q\ge 1$ this
happens if the horizon distance is less than a critical value $\sim
\sqrt{gq}\, m^{-1}_W$ corresponding to a black hole mass
$M_{\rm unstable}(q)\sim \sqrt{gq} \,m_{\rm Pl}^2/m_W $.  (The corresponding
formulas for $q=1/2$ are obtained by replacing $g$ with $g-2$.)  If
one of these unstable solutions is perturbed, it will classically
evolve to some other black hole solution.  Since the total magnetic
flux through the horizon must be conserved and the horizon cannot
bifurcate (at least classically), this must be a solution of the type
we seek, with nontrivial matter fields outside the horizon and the
original magnetic charge.
Thus, there must be new black hole
solutions for all values of the magnetic charge such that the extremal
Reissner-Nordstrom horizon distance is small enough to allow
instability; i.e., for
$$ 1 \le q < q_{\rm cr} \sim e^2g \left({m_{\rm Pl}\over m_W}\right)^2
       \eqn\Qrange$$
and for $q= 1/2$ if $g-2 \ga (m_W/e\, m_{\rm Pl})^2$.
(If $m_W \ga e\sqrt{g}\,m_{\rm Pl}$, the inequalities in Eq.~\Qrange\
cannot be satisfied, and so we do not expect to find new solutions.
\REF\ortiz{M.E.~Ortiz, \pr{45}{2586}92.}
In the context of the spontaneously broken $SU(2)$ theory, it
was shown [\hair,~\ortiz] that the static nonsingular monopole solution is
absent if
the vector boson mass becomes this large.)
For any given charge in this range, there will be new solutions with
masses ranging up to  $M_{\rm unstable}(q)$ (actually, it appears that
the maximum mass is a bit higher,
although of the same order of magnitude) and down to an extremal value
$M_{\rm ext}(q)$.  Without spherical symmetry,
we cannot derive the precise analogues of Eqs.~\extmass\ and \extrad.
However,
we expect the extremal horizon size to scale roughly with $q$, as it
does in the Reissner-Nordstrom case, and so expect
$$  M^{RN}_{\rm ext}(q) \left( 1-{ g^2 \over 4\lambda} \right)^{1/2}  +
      M_{\rm mon} \, \la\, M\, \la \,
      {\sqrt{gq} \, m_{\rm Pl}^2 \over  m_W} + M_{\rm mon} \,.
   \eqno\eq $$

    Matters are somewhat different if $g^2=4\lambda$.  In this case
nonsingular static solutions might be possible and could be found by
minimizing the energy among a class of configurations with fixed
magnetic charge.  However, in the absence of spherical symmetry, the
minimum energy configuration for any integral value of $q$ might
simply be a collection of infinitely separated monopoles of lower
charge.  This is, in fact, what apparently happens in the
spontaneously broken $SU(2)$ theory (except in the
Bogomol'nyi-Prasad-Sommerfield limit).
The existence of solutions with half-integer $q$
depends on whether or not it is possible
to construct a finite energy configuration with such charges (we
can show that this cannot be done for $q=1/2$, but do not have a
result for $q \ge 3/2$).  If this is possible for some half-integer
values of $q$, then there will be a static solution with the lowest
such charge.

    Now consider black hole solutions when $g^2=4\lambda$.  The
Reissner-Nordstrom
solutions are still unstable for small enough masses, but there is no
guarantee that the end point of their classical evolution is a black
hole with the same magnetic charge.  For integer $q$, the classical
instability could eventually lead to a Schwarzschild black hole plus a
number of nonsingular monopoles.  For half-integer $q$, matters are
more complicated.  If $g>2$, the Reissner-Nordstrom solution with
magnetic charge $1/(2e)$ is unstable for small enough mass. Since
there are no nonsingular monopoles with this charge, there must be a
new black hole solution with $q=1/2$, but there need not be any with
$q>3/2$.  If $g \le 2$, the $q=1/2$  Reissner-Nordstrom solution is
stable, but those with $q \ge 3/2$ need not be.  There must be either
a nonsingular monopole or a new black hole solution with $q=3/2$,
although not necessarily both; one cannot conclude anything about the
solutions with higher charge.

     Let us now address the formation and evolution of these new types
of black holes.  In theories with nonsingular monopoles of charge $Q_M
=1/e$ (i.e., those with $g^2=4\lambda$) a Reissner-Nordstrom black
hole whose charge was an integral multiple of $1/e$ could form
by the absorption of magnetic monopoles by an uncharged Schwarzschild
black hole or by the collapse of matter containing magnetic monopoles.
Black holes with half-integer charge could not be formed by these
mechanisms.   Instead,
these would have to be produced in pairs,
perhaps by a quantum tunnelling process in a strong magnetic field.
No matter what the production mechanism,
evaporation via the Hawking process would cause the
black hole mass to decrease and the horizon to shrink.  When the mass
fell below $M_{\rm unstable}$, the black hole would cease to be
Reissner-Nordstrom, and a $W$ cloud would develop outside the horizon.
As evaporation proceeded further and the horizon moved inward,
it would be energetically favorable for nonsingular monopoles to be
emitted.  For a black hole with integer charge, this would continue
until the charge was reduced to $1/e$; in the
final stage of evaporation the horizon would disappear, leaving behind
a nonsingular monopole.  If there are no nonsingular monopoles with
half-integer charge, black holes with half-integer charges would not
evaporate completely; instead, they would eventually evolve to a black
hole of minimal half-integer charge.

     In theories without nonsingular monopoles (i.e., $g^2 < 4
\lambda$), all
magnetically charged black holes would have to be produced in pairs.
The evolution of these objects would be somewhat different.  As in the
previous case, the Hawking process would take a Reissner-Nordstrom
black hole down to $M_{\rm unstable}$, at which point a $W$ cloud would
appear outside the horizon.  With further evaporation the horizon
would continue to gradually contract until it had reached extremal
size.  Because the Hawking temperature of the resulting
extremal hole vanishes, evaporation would cease at this point.
However, further evolution might still be possible.  Magnetic black
holes with masses less than that of the extremal Reissner-Nordstrom
solution have the unusual property that at large separation the
Coulomb repulsion between a pair of holes is stronger than their
gravitation attraction.  (Similar behavior has been noted in theories
with massive dilatons\ref{J.~Horne and G.~Horowitz, \np{399}{169}93.}.)
One could ask whether it
would be possible for a black hole in this mass range to split into
two holes of lower charge.  This process is forbidden classically, but
it might be possible quantum mechanically.

To summarize, we have shown that the existence of finite energy
classical magnetic monopole solutions need not be associated with
nontrivial topology.  The consideration of such solutions leads
naturally to a new class of magnetically charged black holes with
hair.

\refout

\end